\documentclass[prl,superscriptaddress,showpacs,twocolumn]{revtex4}
\usepackage{graphicx}        

\newcommand\VV{\setbox0=\hbox{V}\hbox{\rm V\raise\ht0
  \hbox to0pt{\hss\vbox to0pt{\hbox{v}\vss}}}}
\def\slashchar#1{\setbox0=\hbox{$#1$}           
   \dimen0=\wd0                                 
   \setbox1=\hbox{/} \dimen1=\wd1               
   \ifdim\dimen0>\dimen1                        
      \rlap{\hbox to \dimen0{\hfil/\hfil}}      
      #1                                        
   \else                                        
      \rlap{\hbox to \dimen1{\hfil$#1$\hfil}}   
      /                                         
   \fi}                                         %
\def\be{\begin{equation}}
\def\ee{\end{equation}}
\def\ba{\begin{eqnarray}}
\def\ea{\end{eqnarray}}
\newcommand{\ice}[1]{\relax}

\newcommand{\GeV}{{\rm GeV}}

\newcommand{\nn}{\nonumber}

\newcommand{\al}{\alpha}

\newcommand{\mts}{M_\tau^2}

\begin{document}

\title{Disentangling perturbative and power corrections
in precision tau decay analysis}

\author{D.~S.~Gorbunov}

\affiliation{Institute for Nuclear Research of the Russian
Academy of Sciences, 117312 Moscow, Russia}

\author{A.~A.~Pivovarov}

\affiliation{Institute for Nuclear Research of the Russian
Academy of Sciences, 117312 Moscow, Russia}

\begin{abstract}
Hadronic tau decay precision data are analyzed with account of both 
perturbative and power corrections of high orders within QCD.
It is found that contributions of high order power corrections are
essential for extracting a numerical value for the strange quark mass    
from the data on Cabibbo suppressed tau decays.
We show that with inclusion of new five-loop perturbative
corrections in the analysis the convergence of perturbation theory
remains acceptable only for few low order moments.
We obtain $m_s(M_\tau) =130\pm 27~{\rm MeV}$ 
in agreement with previous estimates.
\end{abstract}

\pacs{12.15.Lk, 13.35.Bv, 14.60.Ef}

\maketitle

Effects of strong interactions is a real stumbling block for investigating 
the electroweak sector of the Standard 
Model~\cite{buras,Kuhn:1998kv,Hollik:ed,Erler:sa}. 
While there remain still many principal problems of QCD as an underlying
theory of strong interactions unresolved, an account of hadronic
effects at the level of few percents is becoming a must 
for the high precision tests of the Standard Model and search for new 
physics~\cite{Erler:2002ix,Davier:2003gq,Kuhn:2003pu,Korner:2003zk,Kuhn:1998ze}.
Although the phenomenon of confinement is still beyond a complete quantitative 
theoretical explanation 
there is a solid qualitative understanding of many features 
of QCD beyond perturbation theory
that allows for a reliable use of perturbation theory (pQCD) 
in its applicability area for obtaining
high precision predictions. The nonperturbative effects are 
accounted for through several phenomenological parameters~\cite{SVZ}.
A high precision achieved for the hadronic $\tau$-lepton decays 
both theoretically and experimentally
makes the $\tau$-system a unique testing ground of 
particle interactions~\cite{schilcher-prl,tauanal,nonpttau}.
The analysis of $\tau$-decays provides information
usable in a variety of ways for: (i) extracting QCD parameters with 
high precision -- strong coupling constant, $s$-quark mass, vacuum
condensates of local operators within the operator product expansion; 
(ii) understanding general properties of perturbation theory and
its asymptotic behavior at high orders;
(iii) evaluating the hadronic contributions necessary 
in the high precision tests of the Standard Model, e.~g. 
the electromagnetic coupling 
$\alpha_{\rm EM}(M_Z)$, muon $g-2$, Higgs mass.

In this note we present a new analysis
of the hadronic $\tau$-lepton decays with the main emphasize on the 
precision and reliability of the theoretical description within QCD.

The total $\tau$-lepton decay rate into tau neutrino and hadrons normalized
to the corresponding pure leptonic decay  
$R_{\tau}=\Gamma(\tau\to h \nu)/\Gamma(\tau\to l\nu\bar \nu)$
splits into a sum of 
strange and non-strange channels
$R_{\tau}=R_{\tau}^{S=0}+R_{\tau}^{S=1}$
with the experimental values
$
R_{\tau}=3.642\pm 0.012$ and  
$
R_{\tau}^{S=1}=0.1625\pm 0.0066
$~\cite{exp12,alephDATA,aleph2002}.
The numerical values for the decay rates $R_{\tau}^{S=0}$
and $R_{\tau}^{S=1}$ are plausibly understood 
since in the parton model approximation
the non-strange and strange parts
of the decay rate are associated with the $ud$ and $us$ decay channels
$ 
R_{\tau}^{S=0}\propto N_c |V_{ud}|^2
$
and 
$
R_{\tau}^{S=1}\propto N_c |V_{us}|^2
$. 
The decay rate is proportional to the number of quark colors in QCD with
$N_c=3$ while the relative difference between the strange and non-strange channels
is due to a numerical smallness of the $V_{us}$ entry of
CKM matrix, $|V_{us}|=0.2196\pm 0.0026$, as compared to the Cabibbo
favored $ud$ channel with $|V_{ud}|=0.9734\pm 0.0008$~\cite{PDG}.
For the determination of detailed characteristics of the
spectrum in $\tau$-decays and further improvements upon precision
the moments of the differential decay rate of the $\tau$ lepton into
hadrons 
\be
\label{defofmom}
R_{\tau}^{kl} 
= \int_0^{\mts} \!\!ds \left(1- \frac{s}{\mts} \right)^k
\left( \frac{s}{\mts} \right)^l 
\frac{d R_{\tau}}{ds}
\ee
have been extracted from the experimental data.
Theoretically, the moments of the differential decay rate $R_{\tau}^{kl}$ are calculable
in QCD perturbation theory (pQCD) within the operator product expansion
for current correlators (OPE)~\cite{cont}. 
The range of indices $(k,l)$ for the moments $R_{\tau}^{kl}$ in
eq.~(\ref{defofmom}) should be properly chosen
in order to guarantee the applicability of QCD perturbation theory for evaluating 
the moments with a strict control over the theoretical precision~\cite{onetwo}.
The analysis of the non-strange part of the decay rate results in an accurate 
determination of the strong interaction coupling constant $\al_s(\mu)$ 
directly in the low energy domain for $\mu\sim M_\tau$~\cite{krajal,bethke}. 
The determination of the numerical value for the $s$-quark mass
exploits the difference
$
\delta \! R_\tau^{kl} 
=R_{\tau S=0}^{kl}/|V_{ud}|^2
-R^{kl}_{\tau S=1}/|V_{us}|^2
$.
This is a sensible setup as CKM is an external quantity to
QCD and should be factor out.
The analysis of the difference $\delta \! R_\tau^{kl}$ 
is more demanding theoretically
and requires much care in the interpretation of perturbation theory
calculations in order to retain the full control over the obtained precision.

The theoretical expression for the difference 
$\delta \! R_\tau^{kl}$ is usually written in the general form 
\be
\label{finaleq}
\delta \! R_\tau^{kl} = 2N_c S_{EW} 
\sum_{n\ge 2}
\delta_n^{kl}(m_s,\al_s)
\ee
where $S_{EW}=1.02$ is the electroweak 
correction~\cite{ewcorrsir}.
The quantities $\delta_n^{kl}(m_s,\al_s)$ with $n\ge 2$ give 
corrections emerging within OPE technique in QCD.
They include mass
corrections within pQCD for $n=2$ and power corrections
for $n>2$. 
Theoretical evaluation of quantities $\delta_n^{kl}(m_s,\al_s)$ 
is very clean as they are computed through the 
two-point correlator of the weak charged currents $j_{\mu}(x)$ 
\ba
\label{correlatorbasic}
i\int dx e^{iqx}
\langle T j_{\mu}(x) j_{\nu}^{\dagger} (0) \rangle
= q_{\mu}q_{\nu} \Pi_q(q^2)+g_{\mu\nu}  \Pi_g(q^2)
\ea
with two scalar form factors $\Pi_q(q^2)$ and $\Pi_g(q^2)$.
Such a decomposition of the general tensor correlator in
eq.~(\ref{correlatorbasic})
into a sum of
scalar form factors avoids kinematical singularities. 
The function $\Pi_g(q^2)$ receives contributions from the
states with the total angular momentum $J=1$ only 
that gives an attractive
opportunity to analyze the decay data with respect to the spin 
content of the particles in the final state~\cite{mirkes}.
The differential $\tau$-lepton 
decay rate is proportional to discontinuities 
$R_{q,g}(s)$ of
the functions $\Pi_{q,g}(q^2)$ across the physical cut along the positive
semiaxis  $q^2=s>0$ in the complex $q^2$ plane
\be
\label{discont}
\frac{d R_{\tau}}{ds}\propto \left(1-\frac{s}{\mts}\right)^2
\left(R_q(s) -\frac{2}{\mts}R_g(s)\right)\, .
\ee
The numerical values for the moments of the $ud$ part of the decay rate 
calculated in QCD are given in Table~\ref{tableUD}. 
\begin{table}[floatfix]
\caption{\label{tableUD}Perturbation theory moments of $ud$ part}
\begin{tabular}{|c||c|c|c|}\hline
$(k,l)$&$(\delta_P)_{kl}^{\rm LO}$ &$(\delta_P)_{kl}^{\rm NLO}$&
$(\delta_P)_{kl}^{\rm NNLO}$\\ \hline
$( 0,0)$&$ 0.200\pm 0.005$ &$ 0.200\pm 0.005 $&$ 0.200\pm  0.005 $\\
$( 1,0)$&$ 0.159\pm 0.004$ &$ 0.167\pm 0.006 $&$ 0.167\pm  0.007 $\\
$( 2,0)$&$ 0.135\pm 0.004$ &$ 0.145\pm 0.006 $&$ 0.151\pm  0.008 $\\
$( 3,0)$&$ 0.120\pm 0.004$ &$ 0.137\pm 0.007 $&$ 0.144\pm  0.009 $\\
$( 4,0)$&$ 0.110\pm 0.004$ &$ 0.125\pm 0.008 $&$ 0.126\pm  0.010 $\\
\hline\end{tabular}
\end{table}
The experimental input for the calculation is a perturbative
correction $\delta_P$ to the total decay rate for the non-strange decays
defined through 
\[
R_\tau^{ud} = N_c S_{EW}|V_{ud}|^2(1+\delta_P+\delta_{NP}+\delta_{EW})
\]
with $\delta_P=0.200\pm 0.005$. The power correction contribution
$\delta_{NP}=-0.003\pm 0.004$ is small and consistent with zero.
The additive electroweak correction is also negligibly small,
$\delta_{EW}=0.001$.
This fixes the numerical value for the moment
$(k,l)=(0,0)$ which is used as the 
input for the determination of the numerical value for the strong
coupling constant $\al_s(M_\tau)$ from $\tau$-decays.
The perturbation theory results are
obtained in the approximation of massless $u,d$ quarks and are rather
stable up to the next-to-next-to-leading order (NNLO), i.e. including two
corrections to the leading order non-vanishing result. 
The inclusion of the fifth order contribution in the coupling constant 
(NNNLO or three perturbation theory corrections to the leading order
non-vanishing result with the ``estimated'' numerical value for the
last coefficient of  perturbation theory expansion $k_3=25$, for
details and discussion
see~\cite{Baikov:2001aa}) 
is still reasonable, we find $(\delta_P)_{10}^{\rm NNNLO}=0.170\pm 0.007$ and 
$(\delta_P)_{30}^{\rm NNNLO}=0.147\pm 0.013$. 
Thus, the pattern of convergence for the perturbation theory
correction to the decay rate is 
\[
(\delta_P)_{10}^{\rm NNNLO}=0.170=0.159+0.008+0.000+0.003
\] 
for the $(1,0)$ moment and
\[
(\delta_P)_{30}^{\rm NNNLO}=0.147=0.120+0.017+0.007+0.003
\] 
for the $(3,0)$ moment.
It is difficult to estimate the actual accuracy of the truncation of
the asymptotic series in the coupling constant. 

The leading power corrections within the operator product expansion
for the correlator from eq.~(\ref{correlatorbasic}) are given by 
$m_s^2$ and quark condensate $m_s\langle \bar s s\rangle$
\ba
\label{correlators}
\Pi_q^{us}(q^2)-\Pi_q^{ud}(q^2)&=&\frac{3}{4\pi^2}\frac{m_s^2}{q^2}
+\frac{m_s\langle \bar s s\rangle}{q^4}+O\left(\frac{1}{q^6}\right) \nn \\
\Pi_g^{us}(q^2)-\Pi_g^{ud}(q^2)&=&
\frac{3}{8\pi^2}m_s^2\ln(\frac{\mu^2}{-q^2})
+O\left(\frac{1}{q^6}\right).
\ea
The correction to $\Pi_{g}(q^2)$ is ultraviolet divergent with 
$\mu$ being a subtraction point.

The $m_s^2$ contribution to $\delta \! R_\tau^{kl}$ reads
\be
\label{masscoefkl}
\delta_2^{kl}(\al_s)=3 F_{kl} \frac{m_s^2}{\mts},\quad
F_{kl}^{\rm LO}=\delta_{l0}+\frac{(k+2)!l!}{(k+l+3)!}\, .
\ee
The mass correction in the function $\Pi_q(q^2)$ from 
eq.~(\ref{correlators}) gives no contribution to the moments with $l>0$
in the leading order of perturbation theory~\cite{Groote:2001im}. 
The leading order result for the $m_s^2$ contribution
from eq.~(\ref{masscoefkl})
gets strongly renormalized in higher orders 
of perturbation theory~\cite{eek20,beta3ref,msNPBPP}. 
In the approach based on the finite order perturbation theory one finds 
\be
\label{f00full}
F_{00}=\frac{4}{3}\left(1+5.333a_s+46.0a_s^2+(283.6+k_2^q)a_s^3\right)
\ee
with $a_s=\al_s(M_\tau)/\pi$. At the $\al_s^3$ order
there
is an unknown constant $k_2^q$ in the correlator $\Pi_q(q^2)$ 
that contributes to the
moments with $l=0$. The use of the moments with $l>0$ allows for pushing
the theoretical accuracy to the $\al_s^3$ level but the experimental 
data for such moments is less precise.
For the numerical value of the strong coupling constant 
$\al_s(M_\tau)=0.344\pm 0.006$ as extracted from the $\tau$ lepton decay 
rate into non-strange hadrons the convergence of perturbation theory
series is slow.
The explicit convergence of higher moments within perturbation theory is worse, e.g.
\be
\label{f20full}
F_{20}=\frac{6}{5}\left(1+6.456 a_s+62.25a_s^2+(547.8+k_2^q) a_s^3\right).
\ee
The detailed analysis of convergence in the finite order perturbation theory
for the moments can be found in refs.~\cite{msNPBPP,onetwo}.
A new momentum to the implementation of perturbation theory results to 
the analysis of $\tau$ physics 
was given by developing the contour improved
method for computation of the moments
that allows to resum the effects of running of the coupling constant
in all orders of perturbation theory~\cite{Pivtau0,groote,DibPich}. 
The technique is especially
transparent in reformulation for the effective scheme description of
the moments~\cite{krajms,effsch}.
Still the explicit convergence of the perturbation theory results 
for the $m_s^2$ correction is slower than for the $ud$ part as 
is seen from Table~\ref{tableUS}. The next order gives
$F_{00}^{\rm NNNLO}=4.011\pm 0.102$ and 
$F_{40}^{\rm NNNLO}=14.108\pm 1.291$ 
with $k_2^q= 160$~\cite{msNPBPP}.
Note that recently the quantity analogous to $k_2^q$ has been computed
for the diagonal vector correlator~\cite{kq2dia}.
The calculational techniques developed for this evaluation (for
details and further references see~\cite{kq2dia}) allows for computing
the quantity $k_2^q$ as well.  
 
\begin{table}[floatfix]
\caption{\label{tableUS}Coefficients of $m_s^2$ term}
\begin{tabular}{|c||c|c|c|}\hline
$(k,l)$&$F_{kl}^{\rm LO}$&$F_{kl}^{\rm NLO}$&$F_{kl}^{\rm NNLO}$ \\ \hline
$( 0,0)$&$1.877\pm 0.009 $&$2.913\pm 0.040 $&$3.475\pm 0.067 $  \\
$( 1,0)$&$1.997\pm 0.016 $&$3.411\pm 0.071 $&$4.408\pm 0.134 $  \\
$( 2,0)$&$2.135\pm 0.023 $&$4.002\pm 0.112 $&$5.613\pm 0.237 $  \\
$( 3,0)$&$2.285\pm 0.031 $&$4.689\pm 0.165 $&$7.152\pm 0.388 $  \\
$( 4,0)$&$2.442\pm 0.040 $&$5.480\pm 0.232 $&$9.109\pm 0.609 $  \\
\hline\end{tabular}
\end{table}

The contribution of quark condensate 
\be
\label{fourpowerkl}
\delta_{4}^{k0}
=-4\pi^2\frac{m_s\langle \bar s s\rangle}{M_\tau^4}(k+2)
\ee
is linear in $m_s$ that implies a potentially large numerical magnitude. 
It suffices to use the leading order approximation of
the coefficient function as the operator 
$m_s \bar s s$ is renormalization group invariant that makes the 
change of the coefficient function small with running.
In the numerical analysis we use
$\langle \bar s s \rangle 
= (0. 8 \pm 0.2 ) \langle \bar u u \rangle $
and
$\langle \bar u u \rangle = - (0.23~\rm{GeV} )^3$~\cite{gammas}.

Further power corrections are written as
\be
\Pi_q=\sum_{n\ge 3}\frac{\langle O^q_{2n}\rangle}{(-q^2)^n},
\quad 
\Pi_g=\sum_{n\ge 3}\frac{\langle O^g_{2n}\rangle}{(-q^2)^{(n-1)}} 
\ee
that leads to $n\ge 3$ contributions
\ba
\label{highpowerkl}
\delta_{2n}^{k0}
=-4\pi^2
\frac{a_{2n}}{M_\tau^{2n}}\frac{(k+2)!}{(n-1)!(k-n+3)!}
\ea
with $a_{2n} =\langle O^q_{2n}\rangle-2\langle O^g_{2n+2}\rangle/\mts$
(see eq.~(\ref{discont})).
Thus, the LO expression for $\delta \! R_\tau^{00}$ reads
\ba
\label{sumrule00}
\frac{1}{2N_c}\delta R_\tau^{00}
=4\frac{m_s^2}{\mts}
-8\pi^2\frac{m_s\langle \bar s s\rangle}{M_\tau^4}
-4\pi^2\frac{a_6}{M_\tau^6}.
\ea
The quantity $a_{6} =\langle O^q_{6}\rangle-2\langle O^g_{8}\rangle/\mts$
contains a contribution from dimension eight 
operators $O^g_{8}$ appearing in the $g$-part of
the correlator~(\ref{correlatorbasic}). Note that this part receives only 
$J=1$ contributions.

The perturbation theory coefficients $F_{kl}$ of the $m_s^2$ contribution
are given in Table~\ref{tableUS}
up to NNLO. An account of new NNNLO results from ref.~\cite{Baikov:2001aa}
shows that the convergence virtually disappears for $k>2$. Thus, 
the $(2,0)$ moment is indeed marginal for the perturbation theory calculations
to be trusted. 
The quantity $\delta R_\tau^{k0}$ is sensitive to the
power corrections up to $a_{2k+6}$ which include the maximal dimension
operator $\langle O^g_{2k+8}\rangle$.

\begin{table}[t]
\caption{\label{tableExp}Experimental moments}
\[
\begin{array}{||c||c|c|c||}
\hline
(k,l)&(0,0)&(1,0)&(2,0)\\ 
\hline
(\delta\! R_\tau^{kl})^{\rm exp}&0.394\pm 0.137&0.383\pm 0.078&0.373\pm 0.054\\
\hline
\end{array}
\]
\end{table}

Experimental results for the moments of the decay rate 
from ref.~\cite{alephDATA} are given in Table~\ref{tableExp}.
Recently published results are rather close~\cite{OPALnew}.
Using these data and theoretical calculation at the next-to-next to
leading order (NNLO) of perturbation theory 
we extract the numerical value for $m_s$. 
With $a_6=0.001~{\rm GeV}^6$ which is obtained in the approximation 
$\langle O^g_{8}\rangle=0$
and neglecting the power corrections $a_{2n}$
with $n>3$ one finds the following results at NNLO:
\ba
\label{nopowerNNLO}
(0,0):\quad m_s(\mts)&=&130 \pm 27_{\rm{exp}} \pm x_{\rm{th}}~{\rm MeV}\nn \\
(1,0):\quad  m_s(\mts)&=&110 \pm 13_{\rm{exp}} \pm x_{\rm{th}}~{\rm MeV}\\
(2,0):\quad  m_s(\mts)&=&{\hskip 2mm}94 \pm{\hskip 1mm}8_{\rm{exp}} 
\pm x_{\rm{th}}~{\rm MeV}\nn 
\ea
Here $x_{\rm{th}}$ denotes a theoretical uncertainty of a given moment
to be discussed in much detail later.
Note that the error in $m_s$ due to the experimental uncertainty of
the measured moments is large. Thus the linear approximation
(propagation of errors) 
is not really applicable for the error analysis of the extracted value
of $m_s$. Considering the leading order of perturbation theory for
the illustration we
obtain the following values for the extracted masses 
\ba
\label{nopowerLO}
(0,0):\quad m_s(\mts)|_{\rm LO}&=&171 \pm 37_{\rm{exp}}~{\rm MeV}\nn \\
(1,0):\quad  m_s(\mts)|_{\rm LO}&=&155 \pm 19_{\rm{exp}}~{\rm MeV}\\
(2,0):\quad  m_s(\mts)|_{\rm LO}&=&141 \pm 13_{\rm{exp}}~{\rm MeV}.\nn 
\ea
For the low value $0.394-0.137=0.257$ 
of the moment $(0,0)$ we find
$m_s(\mts)=171-37=134~{\rm MeV}$ while for the high 
value $0.394+0.137=0.531$ 
we have $m_s(\mts)=171+31=202~{\rm MeV}$.
In our analysis we take the biggest error as a conservative estimate 
for the
uncertainty of the extracted numerical value for the quark mass.

It is instructive to analyze how the theoretical predictions for the 
moments are composed from the different QCD contributions.
To see the relation between perturbation theory and nonperturbative contributions
is most interesting.
We obtain the following decomposition of the numerical values for the
experimental moments into the contributions of perturbation theory 
and power corrections at NNLO:
\ba
\label{nopowerBackNNLO}
(0,0): 0.394&=&0.334 +0.060\nn \\
(1,0): 0.383&=&0.306 +0.077\\
(2,0): 0.373&=&0.286 +0.087\nn 
\ea
The first term is a ``trivial'' perturbation theory power correction
term proportional to the mass squared of the strange quark 
-- $m_s^2$-term. It is explicitly independent of the vacuum structure of QCD or
the numerical values of the vacuum condensates. The second term is
given by the quark condensate and is linear in $m_s$ that makes it
potentially large numerically. This term is renormalization group 
invariant. Note also 
that by construction the ratios of quark condensate 
contributions in different perturbation theory orders are equal to the ratios
of the corresponding values of quark masses. 
One sees that the contribution of the leading power
correction is numerically significant. The relative magnitude of the 
quark condensate contribution 
increases for larger $(k,0)$ moments from 18\% to 30\%.
This calls for the analysis 
of contributions of higher order power corrections.

There are no general systematic techniques to estimate the theoretical errors
due to truncation of the asymptotic series. This is really difficult
problem that should be considered for any given case.
Let us consider a pattern of convergence 
for the extracted masses in different
orders of perturbation theory for coefficient functions:
\ba
\label{patternMass}
(0,0):\quad m_s(\mts)&=&130=171-30-11~{\rm MeV} \nn \\
(1,0):\quad m_s(\mts)&=&110=155-31-14~{\rm MeV}\\
(2,0):\quad m_s(\mts)&=&{\hskip 2mm}94=141-32-15~{\rm MeV}. \nn 
\ea
Taking one half of the last term as the estimate of the truncation
error we find the following results
\ba
\label{thErrorMass}
(0,0):\quad m_s(\mts)&=&130\pm 6~{\rm MeV} \nn \\
(1,0):\quad m_s(\mts)&=&110\pm 7~{\rm MeV} \\
(2,0):\quad m_s(\mts)&=&{\hskip 2mm}94\pm 8~{\rm MeV}.\nn 
\ea
This uncertainty comes from the truncation of the perturbation theory series 
for the coefficients $F_{kl}$.
Combining them with the experimental errors we finally get
\ba
\label{finalEst}
(0,0):\quad m_s(\mts)&=&130\pm 27_{\rm exp}\pm 6_{\rm tr}~{\rm MeV}\nn \\
(1,0):\quad m_s(\mts)&=&110\pm  13_{\rm exp}\pm 7_{\rm tr}~{\rm MeV}\\
(2,0):\quad m_s(\mts)&=&{\hskip 2mm}94\pm {\hskip 1mm}8_{\rm exp}\pm 8_{\rm tr}~{\rm MeV}.\nn 
\ea
For higher moments the relative weight of truncation error is larger. 

Still, this is not a whole story and 
we have to estimate uncertainty due to higher power
corrections.
One possibility to perform such an estimate
is to consider the ``feedback'' series for the moments.
For the $(0,0)$ moment there is no contribution of higher power 
correction terms that leaves it untouched.

For the $(1,0)$ moment there is a contribution
of one higher power correction.
Taking again one half of the last term we get the uncertainty 
$0.077/2\approx 0.04$ from eq.~(\ref{nopowerBackNNLO}).
This uncertainty leads to an additional error in the mass
of $7~{\rm MeV}$ (this number can be obtained either by
a direct computation or from the results for the experimental error 
in the linear approximation, it should be around one half of it,
$13/2\approx 7$).

For the $(2,0)$ moment there are at least two next power
correction contributions. This fact can make the resulting 
error larger if these corrections 
are correlated. Still taking one half of the last term 
we get the uncertainty $0.087/2=0.044$ 
from eq.~\ref{nopowerBackNNLO}.
This uncertainty leads to the additional error in the extracted mass
of $7~{\rm MeV}$. Having in mind the possibility of correlation
of power corrections one could add 50\% to the error
(see also eq.~(\ref{pcFeedbackF}))
and finally get $10~{\rm MeV}$.

This analysis leads to the following uncertainties due to power
corrections
that should be added to the perturbation theory uncertainties from 
eq.~(\ref{thErrorMass}) 
\ba
\label{thErrorMassFin}
(0,0):\quad m_s(\mts)&=&130\pm 6\pm 0~{\rm MeV}\nn \\
(1,0):\quad  m_s(\mts)&=&110\pm 7\pm 7~{\rm MeV}\\
(2,0):\quad  m_s(\mts)&=&{\hskip 2mm}94\pm 8\pm 10~{\rm MeV}.\nn 
\ea
These uncertainties are only indicative as it is 
really difficult to make any reliable evaluation of the errors
due to higher power corrections.
 
A remarkable feature of the obtained results for the numerical value
of the strange quark mass
is the strong dependence of the
extracted numerical value for $m_s(\mts)$ on the particular
moment used for the analysis~\cite{krajms,newpich,Gamiz:2004ar}. 
The theoretical errors are dominated by 
the uncertainty due to the truncation of the perturbation theory series.
The experimental and theoretical errors are independent and added in
quadrature give the total uncertainty of $28$, $16$, $15~{\rm MeV}$ 
for the $m_s$ values extracted from the $(0,0)$, $(1,0)$, $(2,0)$ moments.
From perturbation theory point of view the moment $(0,0)$ is the best one
but its experimental accuracy is low.
With neglecting the contributions due to higher power corrections 
the value of $m_s$ extracted from the $(1,0)$ moment has a 
smaller total uncertainty because the experimental accuracy for the
$(1,0)$ moment is much better than that of the $(0,0)$ moment.
This analysis makes clear that in order to improve upon 
the total precision one should reduce the experimental errors for 
the low moments.

Three estimates for the numerical value of $m_s$ in
eq.~(\ref{finalEst}) are consistent within
the error bars since the experimental errors are rather large. 
However, the situation is
definitely unsatisfactory as there is a systematic
decrease of the central value for $m_s$ extracted from different moments. 
The reason for this systematic
decrease can quite be the neglect of higher power corrections
in the analysis.
The analysis of experimental data with the inclusion of 
higher power corrections as free parameters of the fit gives 
$m_s(\mts)=130 \pm 27_{\rm{exp}} \pm 6_{\rm{th}}~{\rm MeV}$,
$a_8=0.05~\GeV^8$,
$a_{10}=-0.3~\GeV^{10}$.
The value for $m_s$ has not changed since  
the higher power corrections do not contribute to the
$(0,0)$ moment. However the equations for moments $(0,1)$ and $(0,2)$
are satisfied
with the same $m_s$ because of contributions due to 
higher power terms $a_{8,10}$.
It is instructive to see how the moments are composed with an account
of the contributions of power corrections (ordered by the dimensionality)
\ba
\label{pcFeedbackF}
(0,0):~0.394&=&0.341|_{m_s^2}+0.061|_{m_s}-0.008|_{a_6} \\ \nn
(1,0):~0.383&=&0.433|_{m_s^2}+0.092|_{m_s}-0.024|_{a_6} \\ \nn
&& - 0.118|_{a_8} \\ \nn
(2,0):~0.373&=&0.552|_{m_s^2}+0.123|_{m_s}-0.048|_{a_6} \nn\\
&&-0.472|_{a_8} + 0.218|_{a_{10}}.\nn
\ea
There is a sizable contribution from higher power corrections
to the high moments. The numerical values for the high dimensional 
condensates are extracted as fit parameters from the data.
Can they be understood from the present knowledge of the numerical
magnitude of power corrections?
The structure of power corrections for the relevant correlator
is $a_{2n}=C_{2n}m_s\Lambda^{2n-1}$ with $\Lambda$ being a
nonperturbative infrared scale
of QCD and the factor $m_s$ appears as the power corrections to the
difference of the correlators should vanish in
the chiral limit.
For $m_s=130~{\rm MeV}$ and $\Lambda\sim m_\rho=0.77~{\rm GeV}$ one finds
$a_8=0.02C_8$ and $a_{10}=0.012C_{10}$. One expects a fast growth of
coefficients $C_{2n}$ with $n$, e.g.~\cite{Groote:2001im}. 
Assuming a factorial growth
$C_{2n}\sim n!$ one obtains for the ratio
$a_{10}/a_8=5\Lambda^2=3~{\rm GeV}^2$ while 
the values extracted from the data is $6~{\rm GeV}^2$.
This is quite reasonable result given the simplicity of the estimate.
Note that the value of $a_6$ seems to be very small within this logic.
It is known however that there is a huge numerical 
cancellation between 
first several low dimensional power
corrections in the vector and axial channels.
Our method of estimation cannot catch such delicate features
of the physical spectrum.

One can also estimate the magnitude of power corrections
phenomenologically. Assuming for simplicity 
that the continuum contribution basically
cancels in the difference of $ud$ and $us$ channel one is left with the
difference of the resonance contributions.
For the axial part it is given by the difference
\be
\Pi_q^{ud}(q^2)-\Pi_q^{us}(q^2)\sim
\frac{f_{ud}^2}{m_{ud}^2-q^2}-\frac{f_{us}^2}{m_{us}^2-q^2}
\ee
where $m_{ud}^2$ and $m_{us}^2$ are some characteristic scales in the
corresponding channels.
Taking $f_{ud}^2=f_{us}^2=0.6/4\pi^2~{\rm GeV^2}$
and $m_{ud}=m_{a_1}=1.24~{\rm GeV}$, $m_{us}=m_{K_1}=1.4~{\rm GeV}$ 
one finds
$|a_8|=0.06~{\rm GeV}^8$ and $|a_{10}|=0.14~{\rm GeV}^{10}$
which is in a reasonable agreement with the fit.
The contribution of the vector channel to the difference is
much smaller because the masses of corresponding resonances
$\rho(770)$ and $K^*(890)$ are smaller.

It seems hopeless to estimate numerical values for
high dimension contributions from first principles. 
Even if the basis of relevant operators
is identified and the coefficient functions are found 
one remains with the
problem of numerical values for vacuum condensates. 
The factorization hypothesis becomes less
trustworthy with increasing dimension of the condensates.
Still our consideration shows that the fit to data is consistent 
with the estimates based on the structure of hadronic spectra.
The emerging picture is fairly sensible from the physical
point of view. The large $k$-moments give a
high energy resolution of the spectrum as they are saturated with
quite a limited part of the total spectrum; in fact, the large
$k$-moments virtually become the exclusive observables. 
For such almost exclusive observables the influence of high
order power corrections should be large.

Note that the precision of experimental data is not sufficiently good
that allows for a (marginal) fit of all the moments within the error bars 
with only one value of the strange quark mass and without including  
the higher power corrections. We do not discuss this possibility 
as power
corrections are known to be definitely included as free parameters of
the fit.

The obtained numerical value for the strange quark mass 
is compatible with results available in 
the literature~\cite{others,narma,gassLeut,oldmass}.
However the central value is larger than the lattice 
results~\cite{lattice,latticems,Gupta:2003fn}.
We obtain $m_s(2~{\rm GeV}) = 125\pm 28~{\rm MeV}$.

To conclude, our current analysis of the $\tau$-decay data shows that 
the requirement of the applicability of perturbation theory 
strictly limits the allowed range of moments with 
the $(k,l)=(2,0)$ moment being almost marginal since for the higher moments 
the convergence of perturbation theory series is virtually absent. The contributions 
of power corrections are large for the high $k$-moments and
have to be retained that stabilizes the value of the strange quark mass
extracted from the data. 

AAP thanks J.~Gasser and G.~Colangelo for discussions, 
T.~Plehn, M.~Pluemacher, I.~Tkachev for their interest in the work, 
V.A.~Matveev and W.~Hollik for support.
We thank J.~H.~K\"uhn for drawing our attention to new results of OPAL
collaboration, many stimulating discussions, interesting comments 
and strict criticism. 
The work is partially supported by RFBR grant \#~03-02-17177,
by the GPRFNS-2184.2003.2 grants
and VW grant \#~I/77 788.
The work of D.G. was supported in part by
a fellowship of the "Dynasty" foundation
(awarded by the Scientific Council of ICFPM), by the
GPRF grant MK-2788.2003.02, 
and by a fellowship of the ``Russian Science Support
Foundation''.


\begin{thebibliography}{999}
\bibitem{buras}
A.~J.~Buras,
arXiv:hep-ph/9806471 

\bibitem{Kuhn:1998kv}
J.~H.~Kuhn,
{\it Prepared for 29th International Conference on High-Energy Physics
(ICHEP 98), 
Vancouver, British Columbia, Canada, 23-29 Jul 1998}

\bibitem{Hollik:ed}
W.~Hollik,
AIP Conf.\ Proc.\  {\bf 670}, 219 (2003).

\bibitem{Erler:sa}
J.~Erler and P.~Langacker,
Eur.\ Phys.\ J.\ C {\bf 15}, 95 (2000).

\bibitem{Erler:2002ix}
J.~Erler,
AIP Conf.\ Proc.\  {\bf 670}, 227 (2003).

\bibitem{Davier:2003gq}
M.~Davier,
arXiv:hep-ex/0312064.

\bibitem{Kuhn:2003pu}
J.~H.~K\"uhn et al., 
Phys. \ Rev. \ D {\bf 68}, 033018 (2003)

\bibitem{Korner:2003zk}
J.~G.~K\"orner et al., 
Phys.\ Rev.\ Lett.\  {\bf 91}, 192002 (2003);
A.~A.~Pivovarov,
arXiv:hep-ph/0410046.

\bibitem{Kuhn:1998ze}
J.~H.~K\"uhn and M.~Steinhauser,
Phys.\ Lett.\ B {\bf 437}, 425 (1998);
S.~Groote et al.
Phys.\ Lett.\ B {\bf 440}, 375 (1998);
A.~A.~Pivovarov,
Yad.\ Fiz.\ {\bf 65}, 1352 (2002)

\bibitem{SVZ}M.A.~Shifman, A.I.~Vainshtein and V.I.~Zakharov,
Nucl. Phys. B \textbf{147}, 385 (1979);
B.V.~Geshkenbein, B.L.~Ioffe and K.N.~Zyablyuk,
Phys. Rev. D \textbf{64}, 093009 (2001).

\bibitem{schilcher-prl}S.~Groote et al.,
Phys. Rev. Lett. {\bf 79}, 2763 (1997);
A.~A.~Pivovarov,
Yad.\ Fiz.\ {\bf 66}, 1 (2003). 


\bibitem{tauanal} 
K.~Schilcher and M.D.~Tran, Phys. Rev. D \textbf{29}, 570 (1984);
E.~Braaten, Phys. Rev. Lett. \textbf{53}, 1660 (1988);
S.~Narison and A.~Pich, Phys. Lett. B \textbf{211}, 183 (1988);
A.A.~Pivovarov, Nuovo Cim. A {\bf 105}, 813 (1992).

\bibitem{nonpttau} 
E.~Braaten, S.~Narison and A.~Pich, 
Nucl. Phys. B \textbf{373}, 581 (1992).

\bibitem{exp12}
ALEPH collaboration, Z. Phys. C \textbf{76}, 15 (1997);
Eur. Phys. J. C \textbf{4}, 409 (1998);
OPAL collaboration, Eur. Phys. J. C \textbf{7}, 571 (1999).

\bibitem{alephDATA}
R.~Barate {\it et al.}  [ALEPH Collaboration],
Eur.\ Phys.\ J.\ C {\bf 11}, 599 (1999).

\bibitem{aleph2002}
M.~Davier and C.~Z.~Yuan,
eConf {\bf C0209101}, TU06 (2002)
[Nucl.\ Phys.\ Proc.\ Suppl.\  {\bf 123}, 47 (2003)].

\bibitem{PDG}
K. Hagiwara et al., Phys. Rev. D \textbf{66}, 010001 (2002). 

\bibitem{cont}
C.~Bernard et al., 
Phys. Rev. D \textbf{12}, 792 (1975);
E.~Poggio, H.~Quinn and S.~Weinberg,
Phys. Rev. D \textbf{13}, 1958 (1976);
R.~Shankar: Phys. Rev. D \textbf{15}, 755 (1977);
K.G.~Chetyrkin, N.V.~Krasnikov and A.N.~Tavkhelidze,
Phys. Lett. B \textbf{76}, 83 (1978);
N. V.~Krasnikov and A. A.~Pivovarov,
Phys. Lett. B \textbf{112}, 397 (1982).

\bibitem{onetwo}J.~G.~K\"orner, F.~Krajewski and A.~A.~Pivovarov,
Eur. Phys. J. C \textbf{12}, 461 (2000), 
{\it ibid} C \textbf{14}, 123 (2000).

\bibitem{krajal}J.~G.~K\"orner, F.~Krajewski and A.~A.~Pivovarov,
Phys. Rev. D \textbf{63}, 036001 (2001).

\bibitem{bethke}
S.~Bethke, 
Nucl.\ Phys.\ Proc.\ Suppl.\  {\bf 121}, 74 (2003)

\bibitem{ewcorrsir}
W.~J.~Marciano and A.~Sirlin,
Phys. Rev. Lett.  \textbf{61}, 1815 (1988);
E.~Braaten and C.S.~Li, Phys. Rev. D \textbf{42}, 3888 (1990);
J.~Erler,
Rev.\ Mex.\ Fis.\  {\bf 50}, 200 (2004)

\bibitem{mirkes}
J.~H.~K\"uhn and E.~Mirkes, Z. Phys. C {\bf 59}, 525 (1993).

\bibitem{Baikov:2001aa}
P.~A.~Baikov, K.~G.~Chetyrkin and J.~H.~K\"uhn,
Phys. Rev. Lett.  \textbf{88}, 012001 (2002);
Phys.\ Rev.\ D {\bf 67}, 074026 (2003);
Phys.\ Lett.\ B {\bf 559}, 245 (2003);
A.~L.~Kataev and V.~V.~Starshenko, Mod. Phys. Lett. A {\bf 10}, 235 (1995).

\bibitem{Groote:2001im}
A.~A.~Pivovarov,
Phys.\ Atom.\ Nucl.\ {\bf 66}, 724 (2003);
S.~Groote, J.~G.~K\"orner and A.~A.~Pivovarov,
Phys.\ Rev.\ D {\bf 65}, 036001 (2002).

\bibitem{eek20}
S.~G.~Gorishny, A.~L.~Kataev and S.~A.~Larin,
Phys. Lett. B \textbf{259}, 144 (1991);
L.~R.~Surguladze and M.~A.~Samuel,
Phys. Rev. Lett. \textbf{66}, 560 (1991);
K.~G.~Chetyrkin, Phys. Lett. B \textbf{391}, 402 (1997).

\bibitem{beta3ref}
T. van Ritbergen, J.A.M.~Vermaseren and S.A.~Larin,
Phys. Lett. B \textbf{400}, 379 (1997);
K.G.~Chetyrkin, Phys. Lett. B \textbf{404}, 161 (1997);
T.~van Ritbergen, J.A.M.~Vermaseren and S.A.~Larin,
Phys. Lett. B \textbf{405}, 327 (1997).

\bibitem{msNPBPP}
K.~G.~Chetyrkin, J.~H.~K\"uhn and A.~A.~Pivovarov,
Nucl. Phys. B \textbf{533}, 473 (1998);
A.~Pich and J.~Prades, JHEP \textbf{9806}, 013 (1998),
{\it ibid} \textbf{9910}, 004 (1999).

\bibitem{Pivtau0}
A.~A.~Pivovarov, Yad.\ Fiz.\ {\bf 54}, 1114 (1991),
Z. Phys. C {\bf 53}, 461 (1992) 

\bibitem{groote}
S.~Groote, J.G.~K\"orner and A.A.~Pivovarov,
Phys.\ Lett.\ B {\bf 407}, 66 (1997),
Mod. Phys. Lett. A {\bf 13}, 637 (1998).

\bibitem{DibPich}
F.~Le Diberder and A.Pich, 
Phys. Lett. B {\bf 286}, 147 (1992).

\bibitem{krajms}
J.~G.~K\"orner, F.~Krajewski and A.~A.~Pivovarov,
Eur. Phys. J. C \textbf{20}, 259 (2001);
A.~A.~Pivovarov,
Lect. Notes Phys. {\bf 647}, 275 (2004) 

\bibitem{effsch}
G.~Grunberg, Phys. Lett. B \textbf{95}, 70 (1980);
A.~L.~Kataev, N.~V.~Krasnikov and A.~A.~Pivovarov,
Phys.\ Lett.\ B {\bf 107}, 118 (1981),
Nucl.\ Phys.\ B {\bf 198}, 508 (1982), 
Erratum, Nucl.\ Phys.\ B {\bf 490}, 505 (1997) 
A.~Dhar and V.~Gupta, Phys. Rev. D \textbf{29}, 2822 (1984).

\bibitem{kq2dia}
P.~A.~Baikov, K.~G.~Chetyrkin and J.~H.~K\"uhn,
``Vacuum polarization in pQCD: first complete O($\alpha_s^4$) result'',
Talk presented at 7th DESY Workshop on Elementary Particle
Theory: Loops and Legs in Quantum Field Theory, Zinnowitz, Germany,
25-30 April, 2004;
Preprint SFB/CPP-04-30.

\bibitem{gammas}
C.~Becchi et al., 
Z. Phys. C \textbf{8}, 335 (1981); 
A.~A.~Ovchinnikov and A.~A.~Pivovarov, 
Phys. Lett. B \textbf{163}, 231 (1985);
N.V.~Krasnikov and A.A.~Pivovarov, 
Nuovo Cim. A \textbf{81}, 680 (1984);
Y.~Chung et al., Z. Phys. C \textbf{25}, 151 (1984);
H.G.~Dosch and S.~Narison, Phys. Lett. B \textbf{417}, 173 (1998).

\bibitem{OPALnew}
G.~Abbiendi {\it et al.}  [OPAL Collaboration],
Eur.\ Phys.\ J.\ C {\bf 35}, 437 (2004)

\bibitem{newpich}
E.~Gamiz et al., 
JHEP {\bf 0301}, 060 (2003).
\bibitem{Gamiz:2004ar}
E.~Gamiz, M.~Jamin, A.~Pich, J.~Prades and F.~Schwab,
arXiv:hep-ph/0408044.

\bibitem{others}
S.G.~Gorishnii, A.L.~Kataev and S.A.~Larin,
Phys. Lett. B \textbf{135}, 457 (1984);
M.~Jamin and M.~Munz, Z. Phys. C \textbf{66}, 633 (1995);
K.G.~Chetyrkin, D.~Pirjol and K.~Schilcher,
Phys. Lett. B  \textbf{404}, 337 (1997);
K.~Maltman and J.~Kambor,
Phys. Rev. D \textbf{65}, 074013 (2002);
Phys. Lett. B \textbf{517}, 332 (2001);
Nucl. Phys. A \textbf{684}, 348 (2001);
Phys. Rev. D \textbf{62}, 093023 (2000);
S.~Chen et al., 
Eur. Phys. J. C \textbf{22}, 31 (2001);
R.~Gupta and K.~Maltman,
Int.\ J.\ Mod.\ Phys.\ A {\bf 16S1B}, 591 (2001);

\bibitem{narma}
S.~Narison and E.~de Rafael, Phys. Lett. B \textbf{103}, 57 (1981).

\bibitem{gassLeut}
J.~Gasser and H.~Leutwyler, Phys. Rept. \textbf{87}, 77 (1982). 

\bibitem{oldmass}
A.~L.~Kataev, N.~V.~Krasnikov and A.~A.~Pivovarov,
Phys. Lett. B \textbf{123} 93 (1983),
Nuovo Cim. A \textbf{76}, 723 (1983).

\bibitem{lattice}
A.C.~Kalloniatis {\it et al.},
\emph{Lattice Hadron Physics}. Proceedings, 
(Cairns, Australia, July 9-18, 2001).

\bibitem{latticems}
ALPHA and UKQCD Collaboration (J. Garden et al.),
Nucl. Phys. B \textbf{571}, 237 (2000);
ALPHA Collaboration (M. Guagnelli et al.),
Nucl. Phys. B \textbf{560}, 465 (1999). 

\bibitem{Gupta:2003fn}
R.~Gupta,
arXiv:hep-ph/0311033.


\end{thebibliography}
\end{document}